\documentclass[runningheads]{llncs}

\usepackage{graphicx}
\usepackage{listings, listings-rust}
\usepackage{tikz,lipsum,lmodern}
\usepackage[most]{tcolorbox}

\newcommand{\sln}{WICAS}
\begin{document}

\title{Weaving the Cosmos: WASM-Powered Interchain Communication for AI Enabled Smart Contracts}
\titlerunning{\sln{}}
% If the paper title is too long for the running head, you can set
% an abbreviated paper title here
%
\author{Rabimba Karanjai\inst{1}\orcidID{0000-0002-6705-6506} \and
Lei Xu\inst{2} \and
Weidong Shi\inst{1}
\\
\email{rkaranjai@uh.edu}\inst{1},
\email{lxu12@kent.edu}\inst{2},
\email{wshi3@uh.edu}\inst{1}}
\authorrunning{Karanjai et al.}
% First names are abbreviated in the running head.
% If there are more than two authors, 'et al.' is used.
%
\institute{University Of Houston, Houston, USA \and
Kent State University, USA}

\maketitle              % typeset the header of the contribution

\begin{abstract}

In this era, significant transformations in industries and tool utilization are driven by AI/Large Language Models (LLMs) and advancements in Machine Learning. There's a growing emphasis on Machine Learning Operations(MLOps) for managing and deploying these AI models. %along with a focus on distributed inferences. 
Concurrently, the imperative for richer smart contracts and on-chain computation is escalating. Our paper introduces an innovative framework that integrates blockchain technology, particularly the Cosmos SDK, to facilitate on-chain AI inferences. This system, built on WebAssembly (WASM), enables interchain communication and deployment of WASM modules executing AI inferences across multiple blockchain nodes. We critically assess the framework from feasibility, scalability, and model security, with a special focus on its portability and engine-model agnostic deployment. The capability to support AI on-chain may enhance and expand the scope of smart contracts, and as a result enable new use cases and applications.
\end{abstract}
\section{Introduction}
% In the contemporary landscape of technological advancements, substantial progress has been observed in the realm of AI and large language models (LLMs)~\cite{devlin2018bert}, notably the generative pre-trained transformer (GPT) models~\cite{brown2020language} and Google PaLM2~\cite{anil2023palm}. These models have reached a stage where they are increasingly being integrated into practical applications~\cite{https://doi.org/10.48550/arxiv.2302.03202,brown2020language}. A particularly intriguing application of AI and LLMs is in the generation of source code including smart contracts from natural language, a development that holds the promise of transforming the programming process. Examples include ChatGPT and related systems like Github CoPilot~\cite{10.1145/3511861.3511863}, which are capable of translating natural language instructions from programmers into source code in various programming languages. Another area of applications is applying LLMs for modeling and predicting time series data. A major advantage of LLM based time series data prediction over all the traditional approaches is that it can support multi-modal prediction, integrating time series data with other unstructured data types. This new capability may open a new frontier in financial modeling. 

In today's technological era, significant strides have been made in the field of artificial intelligence and large language models (LLMs), such as the advancements seen with GPT models~\cite{brown2020language} and Google's PaLM2~\cite{anil2023palm}. These models are now progressively utilized in real-world applications~\cite{https://doi.org/10.48550/arxiv.2302.03202,brown2020language}. An exciting application of AI and LLMs is their use in generating source code, including smart contracts, directly from natural language descriptions. This application, exemplified by tools like ChatGPT and Github CoPilot~\cite{10.1145/3511861.3511863}, translates programmers' instructions into code across various languages, potentially revolutionizing the programming landscape. Furthermore, LLMs are being applied in forecasting and modeling time series data, outperforming traditional methods by enabling multi-modal predictions that incorporate both time series and other forms of unstructured data. This innovation could greatly enhance financial modeling techniques.

Amidst this backdrop, the increasing use of decentralized ledgers and smart contracts, particularly in the financial sector, is noteworthy. For example, Uniswap's smart contracts managed transactions worth approximately \$7.17 billion daily in 2021~\footnote{https://decrypt.co/63280/uniswap-trading-volume-exploded-7-
billion-heres-why}. Given the growing importance of smart contracts in various contexts, such as Confidential Computing \cite{10.1145/3505253.3505259,10174906}, Decentralized Serverless Functions~\cite{10174945}, and Event-based Transactions \cite{10.1145/3464298.3493401,9461133}, it becomes crucial to investigate whether AI and LLMs can be used effectively for finance modeling and predictions in a distributed and safe way while augmenting a smart contract workflow. 

In this research, we explore the feasibility of executing an AI agent or machine learning model through a smart contract on a blockchain. Our focus is on developing a system that integrates with the current smart contract workflows seamlessly, without necessitating disruptive changes to the existing infrastructure. We aim to adhere to established standards while utilizing AI-driven financial models to produce numerical, textual, and multi-modal outputs. The core of our investigation is to determine whether our proposed system can function within well-established frameworks and offer an innovative approach to interacting with AI models, including LLMs, directly on-chain.

In our study, we introduce a framework designed to enable decentralized on-device AI model agnostic inferences for smart contracts. These smart contracts act as triggers for invoking the AI inference engine via a WebAssembly runtime~\cite{haas2017bringing}. Our research primarily seeks to address the following questions:

\begin{tcolorbox}[colback=red!5!white,colframe=red!75!black]
  \textbf{RQ1: Can smart contracts utilize AI and LLM inferences with reasonable performance?} We try to see if using existing smart contract platforms, we can build a workflow that is AI model and LLM agnostic. 

\textbf{RQ2: Can we get inference locally in a secure way?} We try to run models locally to see if the smart contracts can execute and invoke the models directly on the device and get the inferences back.

\textbf{RQ3: Can we utilize AI accelerators such as GPU for fast inference while executing a smart contract?} Most of the AI and LLM models are compute-intensive and though we can run some of them in CPU, running them using GPU is desirable for reasonable inference throughput. We evaluate if our framework is capable of doing this.

\textbf{RQ4: What are the security implications?} We explore the security implications of such a framework its benefits over traditional methods and pitfalls. Here we show how using our framework inherently mitigates some covert channel security attacks, which are possible for native inferences in the same machine.
\end{tcolorbox}

To address the above, we propose \underline{W}ASM-Powered \underline{I}nterchain \underline{C}ommunication for \underline{A}I Enabled \underline{S}mart Contracts or \sln{} in short. Which uses \cite{kwon2019cosmos} for the blockchain environment to execute a smart contract and uses our reference implementation of AI inferences with open-source models. 

In summary, the contributions of this work are listed below:
\begin{tcolorbox}[colback=green!5!white,colframe=green!75!black]
\begin{itemize}
    \item We propose \sln{} which provides a way to get AI and LLM invocation and response based on a smart contract execution, a new capability for smart contracts that may enable and power emerging application scenarios in the blockchain and Web3 domain. 
    \item We demonstrate the viability of AI and LLM inferences based on smart contracts on local models.
    \item We explore the potential security benefits and considerations for the proposed system.
    \item We demonstrate that our framework is an AI model and underlying inference engine agnostics and can be adapted for other future systems.
\end{itemize}
\end{tcolorbox}

To the best of our knowledge, we are the first to propose and evaluate a system to (1) run AI and machine learning inference based on smart contract execution on a device and (2) show that it is possible to build such a system in heterogeneous systems with varying CPU/GPU and scalability.
\section{Background and Motivation}
The programming models for smart contracts, such as Solidity, face significant limitations in implementing the complex mathematical models that are commonplace in traditional quantitative finance. These limitations are primarily due to constraints that prevent the use of high-performance computing required for sophisticated quantitative models within smart contracts, thereby impeding innovation in Decentralized Finance (DeFi) and other emerging blockchain applications~\cite{shah2023systematic}. Moreover, recent significant advancements in artificial intelligence present new opportunities for blockchain and the decentralized web. The traditional finance sector and the DeFi space are beginning to explore the transformative potential of AI in enhancing financial decision-making, dynamic pricing, prediction, and portfolio optimization~\cite{cao:2020}. Among various AI technologies, large language models (LLMs) and generative AI are viewed as particularly promising for their potential applications in the finance industry~\cite{krause:2023}.
Generative AI consists of a wide range of technologies that can synthesize new data instances based on statistical patterns. Generative AI models targeting finance applications have the potential to revolutionize both traditional financial modeling and decentralized finance by their capabilities to evaluate massive amounts of numerical and textual data, generate time series, and predict financial performance. Besides data-driven modeling in quantitative finance, large language models present promising opportunities for general-purpose on-chain decision-making and governance where natural language content in conjunction with code can be applied for specifying rules and contract logic.

In Cosmos-based blockchains, smart contracts can be executed as WebAssembly (WASM) code, providing a robust environment for implementing smart contracts beyond traditional approaches. Although Cosmos can facilitate Ethereum Virtual Machine (EVM) compatibility through Ethermint, allowing for the porting of existing Solidity contracts, this work concentrates on the built-in support for WASM smart contracts within the Cosmos ecosystem. The Cosmos SDK supports the execution of WASM smart contracts through a dedicated module, x/WASM. Thanks to the modular design of the Cosmos SDK, the WASM module can seamlessly interact with other modules, such as the staking and bank modules. Notably, the EVM support within Cosmos is also modular, implemented in a similar fashion to the WASM module. This modular architecture enhances flexibility and integration possibilities within the Cosmos ecosystem.

% Through messages and queries, the WASM module can access other Cosmos modules for operations such as staking, token transfer, swapping. The messages are encoded. Using plugin, WASM contracts can issue requests and queries to the native modules. The requests and queries are dispatched to the corresponding module interface (called keeper). The WASM interface is flexible enough to support custom  extension. In our case, the goal is to equip WASM contracts with the capability to run high performance computing code in GPU and invoke AI inferences.

\subsection{Transformer Models and Its Applications}

The GPT series of language models, which utilize the Transformer architecture~\cite{vaswani2017attention}, operate in an auto-regressive manner~\cite{brown2020language}. The process starts with a sequence of tokens known as a prompt. The model attentively analyzes this initial input, progressing iteratively. In each iteration, it evaluates the likelihood of possible subsequent tokens. Through intensive training, the model learns the complex process of prediction and selection, thereby refining its ability to generate coherent and contextually appropriate text.

The procedure for processing and sampling to generate a single output token in a language model is known as an iteration. After being trained on a substantial dataset, models like GPT are capable of performing language tasks with notable proficiency. For instance, when presented with the prompt "knowledge is," the model is more likely to predict "power" rather than "apple" as the next word. Following this iteration, the generated token is appended to the initial prompt and re-fed into the model for the generation of the next token. This cycle repeats until an End of Sequence token is emitted, indicating the end of the text, or until a predetermined maximum output length is reached. This inference method, characterized by its iterative nature, contrasts sharply with architectures like ResNet, which typically exhibit a fixed and predictable execution time~\cite{gujarati2020serving}. While each individual iteration of the language model may be predictable, the total number of iterations—and thus the total inference time—can vary.

Additionally, it is important to mention that large language models (LLMs) are not solely confined to language tasks; they can also be easily configured to model and predict time series data.

% \subsection{Transformer Applications}

% At its core, GPT is a word prediction machine for language modeling. However, by feeding it specific instructions (prompts), we can bend it to tackle various NLP tasks. ChatGPT \cite{openai2023Introducing} proves this principle. It's built on top of GPT, but with extra training via supervised learning and human feedback reinforcement (RLHF). This lets ChatGPT handle diverse tasks interactively, from translation to analyzing emotions and creative writing. But this interactivity is demanding. Many users can hit ChatGPT at once, each expecting quick responses. So, how fast it finishes each job or the Job Completion Time becomes critical for keeping ChatGPT-like applications smooth and efficient.

\subsection{Inference Engine}

Inference serving systems like TensorFlow Serving \cite{olston2017tensorflow} and Triton \footnote{https://github.com/triton-inference-
server/server} act as middleware for Deep Neural Networks (DNNs), handling resource allocation, task execution, and result delivery. They maximize hardware utilization, especially on GPUs, by batching multiple tasks. This batching improves overall efficiency but increases memory usage compared to single-task processing. This becomes especially critical for Large Language Models (LLMs) with their heavy memory footprint, limiting their batch size during inferences.

The surge in popularity of GPT models has prompted optimizations in these serving systems to accommodate GPT's unique design and iterative generation process. At the heart of GPT is the Transformer architecture, featuring the distinctive Masked Self-Attention module. This module generates three key values (query, key, and value) for each input token, allowing each token to "see" how relevant other tokens are, regardless of their position. Masked Self-Attention ensures causality by hiding future tokens from each token during prediction, enabling GPT to focus on the next token in the sequence. The attention mechanism essentially grants each token a global view of the entire input sentence, considering the importance of every other token in its prediction.

Within the GPT inference loop, the attention mechanism relies heavily on key-value pairs from preceding tokens. Traditionally, these are recalculated for each step, incurring significant computational overhead. To address this inefficiency, Fairseq \cite{ott2019fairseq} proposes a caching scheme, pre-computing and storing these elements for subsequent use. This two-stage process involves an initial analysis phase where caches are built for each GPT layer based on the prompt. During decoding, only the newly generated token's key, value, and query are computed, with the cache progressively updated. This optimized scheme significantly reduces per-iteration computational workload compared to the initial computations. Similar caching optimizations are also employed by libraries such as HuggingFace\cite{wolf2019huggingface} and FasterTransformer\cite{chelba2020faster}.

Orca\cite{yu2022orca}, on the other hand, introduces a distinct scheduling approach known as iteration-level scheduling. This contrasts with the typical batch processing paradigm by handling just one iteration per batch. While this enables dynamic job entry and completion, it presents challenges in terms of GPU memory limitations and the stringent latency requirements of interactive applications. Therefore, while Orca offers flexibility, it effective utilization requires meticulous resource management.

\subsection{Blockchain}

Blockchain has emerged as a pivotal decentralized framework for data management and storage, serving as the foundation for various distributed applications, including digital currencies\cite{nakamoto2008bitcoin}, decentralized finance,  and networks\cite{luo2023escm}. It functions as a distributed ledger system, enabling transaction recording and verification without relying on central authority or third-party validation. This technology fosters a peer-to-peer network\cite{zyskind2015decentralizing} where participants collaboratively maintain a transparent and immutable public ledger\cite{cao2022blockchain}. Blockchain's structure integrates a physical network that supports communication, computing, and data storage. This infrastructure underpins features like the blockchain consensus mechanism, forming a dual-layer system comprising the physical network and the blockchain. Transactions in blockchain encapsulate client information, recorded in blocks that collectively form a chain, delineating the logical sequence of these transactions. The system's security and efficacy hinge on the consensus mechanism and smart contracts, automating transaction execution and maintaining system integrity. These attributes position blockchain as a critical technology for enhancing AI's reliability and trustworthiness.

\section{Architecture}

For our efforts, we have chosen the Cosmos Network\cite{kwon2019cosmos} as our network of choice for smart contracts. Within the Cosmos network, smart contracts are like programmable gears driving communication and transactions between different blockchains. These contracts act like autonomous robots, following pre-defined instructions without anyone needing to constantly pull the levers. They're crucial for the whole system to work smoothly.

To build and run these robot gears, Cosmos has two handy tools: CosmWASM \footnote{https://github.com/CosmWasm/wasmd} and Ethermint \footnote{https://docs.ethermint.zone}. CosmWASM is like a custom Lego set for developers, letting them build all sorts of different contracts from scratch. Ethermint, on the other hand, acts as an adapter, allowing developers to use code already built for the Ethereum blockchain with Cosmos.

In this paper we use CosmWASM, the native Cosmos smart contract support, as our infrastructure for executing the smart contracts.

\subsection{CosmWASM}

CosmWASM is a specialized platform for building smart contracts within the Cosmos blockchain ecosystem. It uses WebAssembly (WASM) as its engine, allowing developers to write performant and secure contracts. A strength of this is language portability. This "language agnostic" design gives the flexibility and choice. Plus, it runs the contracts in a safe virtual machine sandbox, so everything scales nicely if the project explodes in popularity. That makes CosmWASM ideal for crafting smart contracts that are both speedy and able to handle a ton of activity.

Ethermint acts as a bridge between the Cosmos ecosystem and Ethereum's smart contract playground. It's like a translator that lets existing Ethereum contracts seamlessly join the Cosmos party without needing a complete makeover. This fusion leverages Cosmos's speedy transaction engine, while keeping things modular like the Cosmos SDK, giving developers plenty of building blocks to play with.

The utilization of these smart contract platforms is broad. In decentralized finance (DeFi), both CosmWASM and Ethermint prove invaluable. Ethermint's Ethereum compatibility simplifies the transfer of existing Ethereum DeFi initiatives to Cosmos.

For applications involving non-fungible tokens and tokenization, CosmWASM's robust and secure framework is well-suited. Additionally, in areas like supply chain and logistics management, the automation features of smart contracts become beneficial, with both CosmWASM and Ethermint providing appropriate tools for these sectors.

\subsection{WebAssembly (WASM)}

Introduced by the W3C in 2015, WebAssembly (WASM) is a project aimed at creating a standardized, high-performance, and secure machine-independent bytecode. It achieves safety through distinct, isolated memory regions: the stack, global variables, and a linear memory area, accessed via type-safe instructions for assured memory safety during native code compilation. WASM's architecture, incorporating capability-based security, ensures both performance and security, managing operating system resources like networking and threading through stringent security protocols.

In edge computing, WASM's architecture is crucial for fast, secure processing. By eliminating risky features while supporting C/C++ compatibility, WASM addresses numerous technological challenges. A prime example is the automotive industry's supply chain fragility, where increasing demands for features clash with the impracticality of adding more microprocessor-based control units. WASM enables automotive manufacturers to share physical hardware, reducing microprocessor demand and manufacturing costs by simplifying hardware requirements. This software-centric approach allows manufacturers to focus on advances in automation, infotainment, and safety, without supply chain concerns.

Edge computing, particularly applications requiring autonomous computing and distributed collaboration at network edge, benefits from WASM. These applications require low latency, a challenge for cloud computing. Edge computing, by processing data locally, overcomes issues of network latency and connectivity, ensuring timely data availability.

\subsection{WASMEdge}

WASMEdge facilitates the deployment of WebAssembly\cite{WebAssembly-2024-01-02} (WASM) in edge computing environments. It allows for the integration of serverless functions, which are WASM executable, into various software platforms. Notably, WASMEdge is versatile in its application, being usable at the edge of cloud networks, serving as an API endpoint in a Function as a Service (FaaS) model, operating within Node command line interfaces, and even within embedded systems  \cite{WebAssembly-2024-01-02}.

Artificial Intelligence (AI) advancements, Natural Language Processing (NLP), and machine learning (ML) are important fields of research and use. Before we get into ML implementation with WASMEdge, let's go over some basic knowledge.

TensorFlow Lite runs on WASMEdge and is optimized for embedded systems. It makes on-device machine learning easier. By removing server communication, this method protects data privacy and helps to prevent network latency and connectivity problems.

% \textbf{Understanding the core components:}

% \textit{GraphDef Documents}: These files, which are essential to model data, provide a legible format for your graph's structure to other processes. They come in binary (.pb) and text (.pbtx) formats, with different readability and verbosity levels.

% \textit{Checkpoint Files}: Using no structural data, these files store serialized variables from a TensorFlow network that represent the state of the variables at various training stages.

% \textit{Frozen Graph}: This is made by combining a GraphDef file with the most recent Checkpoint file, then "freezing" (turning variables into constants).

% \textit{TFLite, or TensorFlow Lite}: TFLite is a condensed, memory-efficient variant of TensorFlow that requires less code and dependencies and is optimized for smaller devices. It makes data reading easier by using flat buffers rather than protobufs and does not require object deserialization.

\subsection{WebAssembly Binary}

The interface between WebAssembly (WASM) and the operating system is facilitated by the WebAssembly System Interface (WASI) \cite{clark2019standardizing}. The interface is standardized and resembles POSIX, with an emphasis on cross-platform compatibility and ease of use. Because of this, it can be used in Trusted Execution contexts (TEEs), edge devices, and Internet of Things contexts. WASM applications may access file systems, networks, and other OS services thanks to WASI, which has 46 functionalities. POSIX calls can be effectively converted into WASI calls by languages like C and Rust. 

Additionally, WASI uses a capability-based security paradigm, which establishes a sandbox environment by requiring WASM runtime to approve resource access. By creating an intermediary layer and restricting applications to particular file system components, it improves security.

\begin{figure*}%[h!]
	\centering
	\includegraphics[height=4cm, width=6cm]{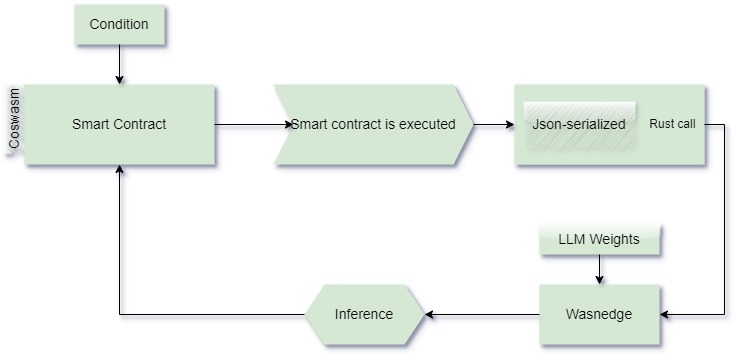}
	\caption{Overview of \sln{}.}   
	\label{fig-overallarchi}
\end{figure*}

For our framework we use WASMEdge since it is the fastest WASM virtual machine available at the moment \cite{zheng2020vm,WebAssembly-2024-01-02}. We explain the high level architecture of \sln{} as shown in Fig. \ref{fig-overallarchi}.

For the contract, we have taken a simple name service sample code from Cosmos SDK and have modified it to pass a prespecified input dynamically created using the name lookup as an input prompt to the AI or LLM inference engine. Here for our AI or LLM inference engine we have chosen WASMEdge. Once the smart contract has been executed successfully, the lookup value is passed as a rust system call to WASMedge.

\section{Implementation Highlights}

For \sln{} we decided to use the fairly large Llama2\cite{touvron2023llama} family of models. Before we can pass the input to the model, the CosmWasm modules has to compile and run the contract.

The compilation process is same for both the smart contract and the Rust inference engine we wrote for loading the model. 

\subsection{Compiling the Contract}\label{compilecontract}
The contract written in Rust has to be compiled to WASM code to be executed. The easiest way for us to do that was to use "cargo". Cargo is Rust's build system and package manager which allows us to easily get packages created in the Rust ecosystem. We use the following to optimize and compile it to a WASM runtime

\begin{lstlisting}[caption={WasmCosm Contract Compile},breaklines, language=bash, label=contractcompile]
RUSTFLAGS= -C link-arg=-s cargo wasm
\end{lstlisting}

Once we have the compiled WASM file, this was deployed in the blockhain.
\begin{lstlisting}[caption={WasmCosm Contract Store},breaklines, language=bash, label=contractcompile]
wasmd tx wasm store nameservice.wasm  --from <your-key> --chain-id <chain-id> --gas auto
\end{lstlisting}

\subsection{Llama 2 Inference}\label{llamainfer}

For our inference engine we chose to use WasmEdge as to use WASM runtime. However to do that we needed a way for us to firs have a program to run LLM inferences (for proof-of-concept and feasibility evaluation). For the inference engine we take inspiration from the excellent llama.cpp \footnote{https://github.com/ggerganov/llama.cpp} project and create a simple Rust program for running the inferences. Here we chose to use Rust as a programming language to remove friction between our smart contract execution on CosmosWasm and here, since our Rust code can directly be integrated into the smart contract and does not need any external message passing scheme to work. 

% \begin{lstlisting}[caption={Wasmedge Code},breaklines, language=bash, label=lst-api-reg]
% wasmedge --dir --nn-preload default:GGML:AUTO:llama-2-7b-chat-q5_k_m.gguf llama-chat.wasm

% [USER]:
% Who is the "father of computer"?
% [ASSISTANT]:
% The term "father of the computer" is often attributed to Charles Babbage, a British mathematician and inventor who conceived the concept of a programmable computer in the 19th century. 
% However, the term has also been used to describe other pioneers in the field of computing, such as Alan Turing, who is known for his work on the Turing machine and the concept of computational universality.
% [USER]:
% Was he the only one?
% [ASSISTANT]:
% No, Charles Babbage was not the only pioneer in the field of computing. He was one of many individuals who contributed to the development of modern computers. 
% .....
% \end{lstlisting}

\subsection{AI Inferences across Nodes}
While our solution works completely fine running in a single node, to be useful in real-world blockchain setting we also explore how it performs if we have multiple Cosmos validator nodes running inferences (use the same AI model). 
One of the challenging use cases of such a setting will be multiple node running an inference with the same machine learning model and an identical input. 

To achieve consensus, it requires AI inference reproducibility.  Since LLMs are not-deterministic, we need a way to support determinism for this use case. Though inference reproducibility is a research topic of its own, existing results in the literature are enough for the purpose of our research. 

Running advanced ML and LLM inference across multiple blockchain nodes could introduce variability in inference results due to several factors. These factors include using deterministic AI code, floating number resolution, node hardware capabilities, and the stochastic nature of some LLMs if they utilize mechanisms like dropout during inference for regularization purposes. It is crucial to understand that blockchain architectures are inherently designed to ensure consensus despite potential discrepancies across nodes. Typically, mechanisms like Proof of Work (PoW) or Proof of Stake (PoS) are employed to achieve agreement among nodes regarding the validity of transactions, which could analogously be applied to agreeing on LLM inference results. 

Based on experiments, the best practice to achieve inference reproducibility across nodes, is to use deterministic implementation of machine learning algorithm, manage random seeds with on-chain support like using on-chain random beacon (a well studied problem in the recent years), ran AI models using 64 bit floating numbers.  It is important to highlight that inference reproducibility has been successfully demonstrated in one application use case where LLMs are used as a covert communication channel between a sender and a receiver. In ~\cite{bauercodaspy}, the authors demonstrate how LLMs can be configured to produce consistent LLM outputs across devices. Our case directly benefits from the findings and best practice identified in the LLM covert communication research for achieving reproducible inferences across validator nodes.  

If the LLM inferences do vary across nodes, the blockchain protocol would need to specify a method for selecting the 'correct' or 'agreed-upon' inference. This is based on a majority rule where the most frequently generated result is chosen, or more sophisticated approaches like weighted scoring based on node reputability or stake can be employed.

%Furthermore, the scalability and performance implications must be considered. As more nodes participate in the inference process, the computational and communication overhead could increase significantly, potentially slowing down the inference process. Techniques such as sharding, where the blockchain is divided into smaller, more manageable segments (shards), could be leveraged to maintain efficiency.

%While running LLM inference in a blockchain framework on a single node does not fully harness the distributed nature of blockchain technology, expanding this to multiple nodes introduces complexities in ensuring consistent and reliable inference across the network. Bauer et al~\cite{bauercodaspy} shows how these can be used to produce consistent LLM outputs.

\section{Portability}
One of the key benefits to our proposed solution is model portability and inference engine (as well as model) agnostic nature. Even though we had chosen WasmEdge to showcase our proposed framework, this can easily be replaced by any other code that can communicate directly with a Web Assembly runtime. The primary reason we chose WasmEdge or any Web Assembly runtime because of their ability to communicate with the WebGPU~\cite{kenwright2022introduction} API. We have also tested our implementation using the well documented WGPU ~\cite{wgpu} API.

\subsection{Inference Engine Portability}

Rust is used to construct the demo inference code, which is compiled to WebAssembly. This core Rust code is only forty lines long and is surprisingly small. It can process inputs from the user, log the flow of the conversation, modify the text to make it compatible with the llama2 input template, and do inference tasks using the WASI NN API. This simplified method demonstrates how well the code manages these operations and how efficient it is.

\begin{lstlisting}[caption={Rust Inference}, style=boxed, breaklines, language=Rust, label=infer rust, basicstyle=\tiny]
fn main() {
    let arguments: Vec<String> = env::args().collect();
    let model_identifier: &str = &arguments[1];

    let neural_network =
        wasi_nn::GraphBuilder::new(wasi_nn::GraphEncoding::Ggml, wasi_nn::ExecutionTarget::AUTO)
            .assemble_from_cache(model_identifier)
            .expect("Successful build");
    let mut execution_context = neural_network.create_execution_context().expect("Context initialization");

.......
.......
.......

        // Input processing.
        let input_tensor_data = combined_prompt.as_bytes().to_vec();
        execution_context
            .assign_input(0, wasi_nn::TensorType::U8, &[1], &input_tensor_data)
            .expect("Input set");

        // Inference execution.
        execution_context.execute().expect("Successful computation");

        // Output retrieval.
        let mut result_buffer = vec![0u8; 1000];
        let result_size = execution_context.obtain_output(0, &mut result_buffer).expect("Output retrieval");
        let response = String::from_utf8_lossy(&result_buffer[..result_size]).to_string();
        println!("Response:\n{}", response.trim());

        historical_prompt.push_str(&format!("{} {}", combined_prompt, response.trim()));
    }
}
\end{lstlisting}

This can be compiled using

\begin{lstlisting}[caption={Rust Compile},breaklines, language=bash, label=rust compile]
curl --proto '=https' --tlsv1.2 -sSf https://sh.rustup.rs | sh
rustup target add wasm32-wasi
\end{lstlisting}

Our handwritten Rust code is also definitely not a requirement. To showcase portability we had directly taken the llama2.c~\cite{karpathy} engine and compiled it into WASM file.

\begin{lstlisting}[caption={llama2.c wasm},breaklines, language=bash, label=llama2.c wasm]
$ run.c -D_WASI_EMULATED_MMAN -lwasi-emulated-mman -D_WASI_EMULATED_PROCESS_CLOCKS -lwasi-emulated-process-clocks -o run.wasm
\end{lstlisting}

\subsection{Model Portability}

Even though we have primarily targeted the Llama 2 family of models. We are not bounded by them. Since we base our implementation on the llama.cpp, the model formats have to be ggml. For which we utilize the GGML plugin for WasmEdge~\cite{WasmEdge-WASINN}. We also have tested the framework with LLAVA~\cite{liu2023llava} which is a multi-modal model. However our inference code is not equipped to handle non-text inputs, hence we only tested the text input and output of the model. However, a base64 encoder and decoder is certainly possible to pass multi-modal (image, numbers, text) input to the model to get its feedback.

\begin{figure*}%[!htbp]
	\centering
	\includegraphics[height=4.2cm, width=7cm]{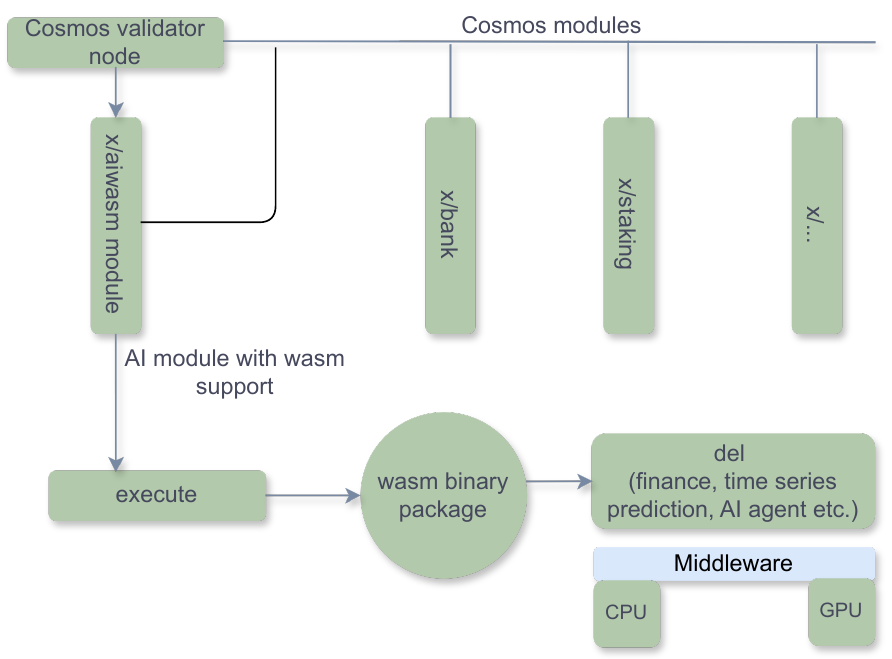}
	\caption{Working of \sln{}.}   
	\label{fig-archi}
\end{figure*}

\sln{} works as shown in Fig  \ref{fig-archi}.

The validator node running in Cosmwasm generates the input message to be passed to the x/aiwasm runtime (which in this case is WasmEdge) which runs the inferences. 

WasmEdge can utilize both GPU or CPU based on what is available in the host system. It uses WebGPU API to see if there is a compatible GPU available, and runs the inference if it is available. If not then tries to run the inference in CPU. \sln{} is compatible with other Cosmos modules since it does not modify anything within the CosmWasm, but only individual contract. 
\section{Discussion}
We have run the LLM/AI execution and Cosmwasm in a machine equipped with Intel I9 Processor and NVIDIA 4080 Mobile graphics processor with 16GB of RAM. We intentionally chose a moderately configured device to note any potential limitation of the system. In our testing both of the models ran in reasonable amount of time. The llama2 model was able to generate a 35 token per second and llava 30 token per second.  

\subsection{CosmWasm and WASM Contract}
We had modified the name space example contract from the Cosmos example repository in Github to incorporate our Rust code. And we were able to run the compiled WASM smart contract as described in Section \ref{compilecontract}.

\begin{tcolorbox}
\textbf{RQ1} \textit{Can smart contracts utilize LLM and AI inferences with reasonable performance?}

We were able to write a proof of concept code and call a LLM/AI inference engine we to produce a model output in a moderately configured commodity hardware. Within our limited scope smart contracts, or more specifically smart contract executed using CosmWasm is able to generate LLM inferences using our framework. Note that LLMs are more demanding in computing resources than other AI models. 
\end{tcolorbox}

\subsection{Security}

Since the input is executed directly from the contract and the contract is compiled into a WASM file before running. The data should be safe as long as the host machine is safe. Also since the inference is being run in either that machine or in another clustered machine, the models and inferences on device and protected as long as the system integrity is protected.

\begin{tcolorbox}
\textbf{RQ2} \textit{Can we get inference in local in secure way?}

The inference is done in local since the compiled models are also in local. It assumes that all the validator nodes have access to the same models.
\end{tcolorbox}

\subsection{Model Performance}

Since we use WasmEdge for our model evaluation, we are able to utilize WebGPU API completely. That gives us access to GPU in the host system if its available, or switch to a CPU runtime based on the inference engine. 

\begin{tcolorbox}
\textbf{RQ3} \textit{Can we utilize GPU for fast inferences while executing a smart contract?}

WebGPU allows us to access and utilize the GPU runtime to run our models in the GPU in an efficient way.
\end{tcolorbox}

\subsection{Security Consideration}

Even though the model is running locally, as the model is loaded into memory, attacks are possible. We specifically look into one specific family of attack that makes most of the local inference system today vulnerable if using native solution.

The VU\#446598~\footnote{https://kb.cert.org/vuls/id/446598} dubbed as \textit{LeftoverLocals} affects all Apple, AMD, Qualcomm family of GPUs. 

% Originally designed for enhancing graphics processing, GPUs have evolved rapidly, emphasizing performance. This rapid development often overshadowed security concerns, which rarely impacted applications significantly. As a result, both GPU hardware and software experienced swift and substantial changes in architecture and programming models. This fast-paced evolution led to intricate system stacks and ambiguous specifications. For instance, in contrast to the extensive documentation available for CPU ISAs, NVIDIA offers minimal guidance, mostly limited to concise tables. Such lack of detailed specifications has been a factor in worrying issues, historically and presently, as demonstrated by cases like LeftoverLocals.

In the context of GPU computing, it has been identified that one GPU kernel is capable of accessing memory values from another kernel, even when these kernels are segregated across different applications, processes, or user environments. The memory area where this interaction occurs is known as local memory. This local memory functions akin to a software-managed cache, comparable to the L1 cache found in CPUs. The size of this local memory varies between GPUs, ranging from tens of kilobytes to several megabytes. Trail of Bits demonstrated this vulnerability's presence~\footnote{https://blog.trailofbits.com/2024/01/16/leftoverlocals-listening-to-
llm-responses-through-leaked-gpu-local-memory/} through various programming interfaces, including Metal, Vulkan, and OpenCL, across diverse combinations of operating systems and drivers. This finding highlights a significant security concern in GPU architecture.

However their findings report that even in vulnerable machines, the models when accessed through WebGPU did not dump any secrets apart from zeros~\cite{trailofbitsLeftover}. This makes \sln{} resilient to these type of memory dump attacks. However, these kind of attacks show us more work is needed to continually protect the system from future threat attacks, even though the present one did not pose a threat.

In case model confidentiality is a concern, AI models can be executed by enclaves and TEEs (trusted execution environments). TEEs such as Intel SGX are capable of running inferences for machine learning models like DNNs. The new GPU TEEs may soon support running LLM inferences in concealed environment. 

\begin{tcolorbox}
\textbf{RQ4} \textit{What are the security implications?}

Attacks like LeftoverLocals are mitigated in this framework. but future work is necessary to understand the security and privacy needs of on-chain AI inferences and assess the attack surface.
\end{tcolorbox}

\subsection{Potential Applications}
The capability of running AI inference engine as part of contract execution could open new opportunities for application areas such as DeFi, decentralized insurance, DAO governance, voting, prediction market, AI agent based decision making, to name just few.

\subsubsection{Open Research Avenues}

% \textbf{Natural language text or multi-modal data as smart contract inputs and outputs}: Ability to understand natural languages allows smart contracts to accept diverse data as inputs. The input can be interpreted by the AI models before it is further processed. 

\textbf{AI based DeFi models}: With the kind of support described in this work, machine learning based finance models can be incorporated on-chain as part of a DeFi contract. This will certainly expand the scope of DeFi to a new level where DeFi contracts can directly apply machine learning on-chain in decision making such as pricing, liquidity optimization, risk reduction. 

\textbf{On-chain governance based on natural language texts}: With LLM support, the concept "code is law" can be expanded to use both code and nature language to regulate the behavior of smart contracts. Complex business logic could be described in natural language and enforced as smart contracts. 

\textbf{On-chain AI agents}: With integration of AI into smart contracts, smart contracts can act as AI agents. This may redefine to our current understanding of smart contracts and significantly increase the scope of smart contracts in future. 

%\textbf{Decentralized AI Marketplaces}: The ability to execute AI models on-chain could lead to the development of decentralized marketplaces where AI models are bought, sold, and executed in a trustless manner.
%\textbf{AI-Driven Smart Contract Automation}: AI models could be used to automate complex decision-making processes within smart contracts, potentially revolutionizing areas like decentralized finance and supply chain management.
%\textbf{Privacy-Preserving AI on the Blockchain}: The paper touches on the potential for privacy-aware AI inference. This could be a significant research area, exploring how to leverage blockchain's security features to protect sensitive data used in AI models.

\subsection{Remarks}

Depending on the types of blockchain, for instance, permissioned or permissionless, running AI inferences may require gas fee. It is outside the scope of this work to specify a detailed gas model for AI inferences. It is plausible to compute gas fee based on model size and the number of LLM tokens (similar to how fee is calculated by OpenAI for using ChapGPT).  In case of Cosmos module, aiwasm module can charge gas fee using a specific fee model implemented by the module. A simple approach is to apply a fee model similar to EVM pre-compiles. Another remark is that although the current work focuses on enabling AI for WASM based smart contracts, it is plausible to apply the same concept to EVM. This could be achieved using dedicated pre-compiles. Last but not the least, it is not assumed that AI inferences must be executed on the main chain. The framework could be applied to support AI modeling in side-chain, application specific chain (app chain), or layer 2 network.

\section{Future Work}

\subsection{WebGPU Native}

Having received universal support from top browser developers, WebGPU~\footnote{https://developer.chrome.com/blog/webgpu-io2023} is an impressive step forward. %It performs far better on GPU efficiency than WebGL. Among the prominent features of WebGPU are storage textures and compute shaders, which allow direct storage in shaders without requiring texture sampling. This lessens the need for data repackaging and makes structured data easier to use. But for our use case WebGPU also gives significant performance improvement and near native speedup for using GPU~\cite{Unlock1}.
There have been some work which tries to run AI inferences directly using WebGPU like tokenhawk~\footnote{https://github.com/kayvr/token-hawk} which tries to implement models by hand, which is not scalable. The other more mature methods all use WASM as a backend like mlc-llm~\cite{MLCLLMHome-2024-01-18} and WebONNX~\cite{webonnx8}. 

\subsection{Usage of WebAssembly (WASM)}

The large language model (LLM) and AI inference engine in our architecture run on both CPUs and GPUs and is written in Rust. This dual compatibility addresses common limitations in web-based computation by facilitating a balance between memory consumption, data transfer volume, and CPU/GPU utilization. %WebAssembly improves this by lowering JavaScript-related overhead by enabling pre-compiled assembly code to run in web browsers. This also allows us to use the same API to access it using Rust. Additionally, WebAssembly facilitates the compilation of a single WASM file from both Rust and C++ (in case of our Rust code and llama2.cpp). It greatly simplifies the programming logic for us. 
However we still need to load the model every time. That is not suitable for large models and specially for execution on-chain and from a smart contract. %It does not allow for native data storing for the inference results.

In our future work, we want to explore these two venues, of (a) coming up with a WebGPU framework that can run inferences in native speed in a programming language agnostic way, communicated just by message passing. Similar to what WebGPU allows us but not limited to Rust. We also want to explore if (b) we can develop a better way to \textit{stream} model weights instead of loading the model completely. That will allow us to have more control for performance optimization. %what gets stored and executed directly from the smart contracts, allowing us to have usage of contracts in spaces like building AI agents on blockchain, insurance negotiators and voting negotiating agents based on smart contracts.
\section{Conclusion}
In conclusion, our research addresses vital questions regarding the integration of AI and Large Language Models (LLMs) with smart contracts. We demonstrated through a proof of concept that smart contracts, particularly those executed using CosmWasm, can effectively generate AI/LLM inferences. %The local execution of these inferences ensures privacy compliance, and 
The use of WebGPU technology allows for efficient GPU-accelerated inferences within a smart contract environment. Our findings affirm the feasibility, performance efficiency, and security of utilizing AI models in blockchain environments. This opens up new avenues for blockchain applications, leveraging the power of AI and LLMs while maintaining security standards.

%
% ---- Bibliography ----
%
% BibTeX users should specify bibliography style 'splncs04'.
% References will then be sorted and formatted in the correct style.
%
% \clearpage
\bibliographystyle{splncs04}
\bibliography{ref}

\end{document}